\documentclass[aip,apl,reprint,superscriptaddress,superbib]{revtex4-1}  

\long\def\symbolfootnote[#1]#2{\begingroup%
\def\thefootnote{\fnsymbol{footnote}}\footnote[#1]{#2}\endgroup}

\usepackage{graphicx}
\usepackage[nice]{nicefrac}
\usepackage[loose]{units}
\usepackage{natbib}

\begin{document}

\title{Coupled-Plasmon-Controlled Transmission in Distributed Bragg Structures}

\author{Charles Rohde}
\affiliation{The Photon Factory, University of Auckland, Auckland, New Zealand}
\email{c.rohde@auckland.ac.nz}

\author{Miriam Deutsch}
\affiliation{Department of Physics, University of Oregon, Eugene, Oregon, 97403, USA}
\email{miriamd@uoregon.edu}

\begin{abstract}
Using the finite element method we investigate plasmon-mediated transmission in a periodically modulated metal-insulator-metal (MIM) grating. We compute the eigenmodes of a silver-silica-silver conformal coating atop an array of close-packed silica rods, and correlate them with extinction and transmission characteristics of the structure. We observe efficient coupling of impinging plane-waves to gap plasmon modes, allowing control of both bandwidth and intensity of the transmitted radiation.
\end{abstract}

\maketitle

\label{sec:intro}

\begin{figure}[b]
\begin{centering}
\includegraphics[width=85mm]{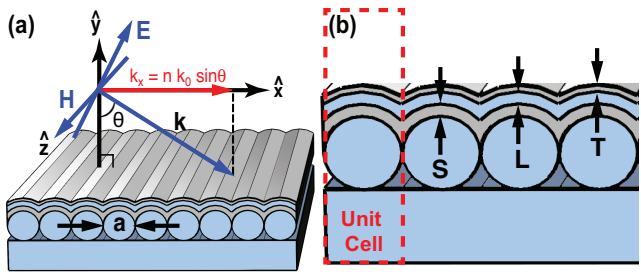}
\caption{Schematic of MIM structure: (a) Coordinate system for a TM plane wave, incident in the $x-y$ plane from vacuum on a MIM grating. (b) Structural parameters: $S=30\unit{nm}$, $L=100\unit{nm}$, $T=15\unit{nm}$. The colors blue (silica) and gray (silver) indicate material composition.}
\label{fig:gcart}
\end{centering}
\end{figure}

Plasmon mediated extraordinary optical transmission relates to enhanced diffractive transmission of light through an optically thick metal film patterned to support surface electromagnetic (EM) modes. This has been first observed in the transmission of light through sub-wavelength hole arrays in otherwise opaque metal films,\cite{Ebbesen:1998p2260} as well as through solid thin metal films with sub-wavelength surface corrugation.\cite{Schroter:1998p3516} The near-field energy confinement and the related surface EM field enhancements associated with surface plasmon polaritons (SPPs) are intimately linked to this effect.

Controlled light transmittance in resonant systems may alternately be achieved by coherent coupling of modes, for example by converting a bright (\emph{i.e.} absorptive) mode into a dark (and therefore nearly transparent) state, as is commonly done in quantum systems using electromagnetically induced transparency (EIT).\cite{Harris:1990p1107} In classical EM systems similar behavior may be obtained through dispersion-engineering. When properly implemented, taking into consideration both materials' and structural resonator parameters, it is possible to realize artificial composite systems with custom dispersions, mimicking the EM response seen in EIT.\cite{Smith:2004p063804, Liu:2009p758}

We have previously shown how concentric metal-insulator shells surrounding a micron-scale metal sphere (MIM resonators) may be designed to exhibit dramatic suppression of their forward-scattering cross sections, resulting in a tunable transparency window.\cite{Rohde:2007p415} From a mode-coupling perspective, this phenomenon is not limited to isolated MIM spheres. Expected to persist in periodic MIMs, the effect may be utilized in applications such as opto-microfluidic sensors and plasmon-enhanced coherent light emission. In this Letter we use finite element modeling (FEM) to demonstrate coupled-plasmon-controlled transmission in MIM-coated cylindrical arrays.

Figure~\ref{fig:gcart} illustrates the geometry, excitation scheme, and computational domain for the system. A MIM grating is formed from the hemi-cylindrical modulation of a MIM film by a close-packed array of silica rods of diameter $a$ resting on a semi-infinite silica substrate. The layers are characterized by their material composition (silver-silica-silver) and respective thicknesses ($S$, $L$ and $T$). Using FEM we calculate the structure's dispersion relations as well as the transmittance, $\mathcal{T}$, and reflectance, $\mathcal{R}$. The incident field, impinging at an angle $\theta$ with respect to the substrate surface normal, is a transverse-magnetic (TM) plane-wave with free space wavevector $\textbf{k}_0$.

In our 2-dimentional FEM simulation, a single unit cell is modeled with periodic boundary conditions along the vertical dashed lines, as shown in Fig.~\ref{fig:gcart}(b). Open boundaries are simulated above (in vacuum) and below the structure (in silica) using the perfectly matched layer technique. The index of refraction used for all silica regions was $n_d=1.42$ and the radius of the rods was chosen as $a=600\unit{nm}$. The cylinders are conformally coated with concentric MIM partial shells as shown. We have used a standard Drude model to characterized the dispersion of the silver film, where the frequency-dependent permittivity $\varepsilon(\omega)$ is given by

\begin{eqnarray}
\varepsilon(\omega) = \varepsilon_r - i \frac{\sigma(\omega)}{\varepsilon_0 \omega}.
\end{eqnarray}

\noindent Here $\varepsilon_r = 4.1$ describes the response to bound electrons and $\varepsilon_0$ is the permittivity of free space. The frequency-dependent conductivity is given by

\begin{eqnarray}
\sigma(\omega) = \frac{\sigma_{dc}}{1+i \tau \omega}
\end{eqnarray}

\noindent where $\sigma_{dc}=62\times10^6\unitfrac{S}{m}$ is the DC conductivity of silver and $\tau=4\times 10^{-14}\unit{s}$ is the electron relaxation time.\cite{ashcroft1976solid}

The dimensions of the silver layers were set thin enough to allow significant SPP coupling across the silica film, while keeping them sufficiently thick to support well defined plasmon eigenmodes. The size of the silica spacer, $L=100\unit{nm}$, was chosen to be small enough to constrain SPP excitations in the dielectric gap to a single symmetric TM mode.\cite{Dionne:2006p3531} EM coupling of the two internal metal-insulator interfaces induces mode splitting of the gap plasmon, resulting in a symmetric and antisymmetric pair. The antisymmetric TM mode is cut off\cite{Avrutsky:2007p3582} when $L\lesssim\lambda/(2 n_d)$, which for $L=100\unit{nm}$ implies that only the symmetric gap plasmon is excited at visible frequencies. Allowing only a single gap mode simplifies the interpretation of the results below.

\begin{figure}
\begin{center}
\includegraphics[width=85mm]{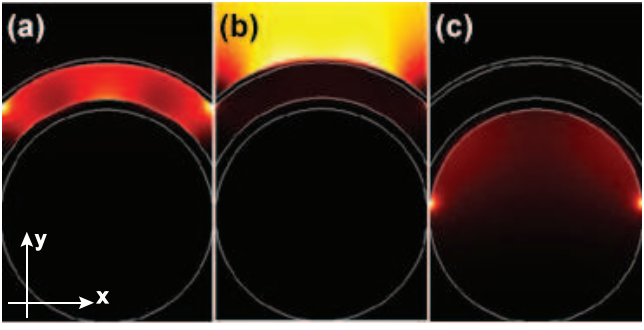}
\end{center}
\caption{EM energy density plots of (a) ISP mode ($\kappa=0$, $\omega_\kappa=1.54\times 10^{15}\unitfrac{rad}{s}$), (b) ESP mode ($\kappa=0.5$, $\omega_\kappa=1.57\times 10^{15}\unitfrac{rad}{s}$), (c) LSP mode ($\kappa=0$,
$\omega_\kappa=1.97\times 10^{15}\unitfrac{rad}{s}$).}
\label{fig:modes}
\end{figure}

The weak form of the TM Helmholtz eigenvalue equation was implemented with the commercially available PDE solver, COMSOL Multiphysics$^{\textregistered}$. In principle the Bloch-vector 
parameterized solution of the quadratic eigenvalue equation for the eigenfrequency $\omega_\kappa$ yields the full EM eigenmode structure and periodic plasmonic band structure. However, the strongly dispersive nature of lossy metallic structures leads to a nonlinear Helmholtz eigenvalue equation for the magnetic field ${\bf H}=H_z{\bf \hat{z}}$:\cite{Fietz:2011p3517}

\begin{equation}
\nabla\times ( \varepsilon^{-1}(\bf{x},\omega_\kappa) (\nabla\times \bf{H}) ) -  \omega_\kappa^2 \bf{H} = 0
\end{equation}
Such nonlinear problems can be solved using an iterative, damped, Newton-Gauss minimization with the addition of a global normalization condition on the eigenvector $\bf{H}$:
\begin{equation}
\int_\Omega H_z(\omega_\kappa) H_z^*(\omega_\kappa)d\Omega / A = 1
\end{equation}
where $A$ is the area of the computational domain $\Omega$.\cite{Anselone:1968p3598}

As is typical with nonlinear iterative approaches, solution convergence is not guaranteed. Solving the above nonlinear Helmholtz equation has been treated in several ways by other authors.\cite{Davanco:2007p3526,Parisi:2012p3525}
In our approach we maximize the likelihood of convergence through careful initialization of the Newton-Gauss method. To find a good approximation of the true eigenvalue/eigenmode pairs at the Brillouin zone boundaries, we solve the eigenvalue problem, equation (3), with a large set of fixed test frequencies from $\omega'_\kappa = 1 \times 10^{15}\unitfrac{rad}{s}$ to $=3 \times 10^{15}\unitfrac{rad}{s}$ for the two parameters $\kappa=0$ and $\kappa=\pi/a$. Solutions which converge are then iterated upon with updated solution frequencies, yielding the set of true eigenvalue/eigenvector ($\omega_\kappa, {\bf H}$)  pairs for the Brillouin zone boundaries. These computationally expensive boundary solutions are then used to initialize the Newton-Gauss method. The periodic structure's band structure is found through the successive use of previous solutions to initialize the Newton-Gauss method as the parameter $\kappa$ is monotonically increased from $\kappa=0$, and also decreased from $\kappa=\pi/a$, for each unique boundary eigenfrequency.

Three types of eigenmode field distributions are found in this system, as shown in Fig.~\ref{fig:modes}.
The general character of the modes can be described as: (i) gap, or interior surface plasmons (ISP), confined to the dielectric layer between the two metal films, (ii) exterior surface plasmons (ESP), associated with excitations along the outermost vacuum-metal interfaces, (iii) localized surface plasmon (LSP) modes associated with the excitation of plasmons on the cusps formed along the contact lines between rods.

\begin{figure}
\includegraphics[width=85mm]{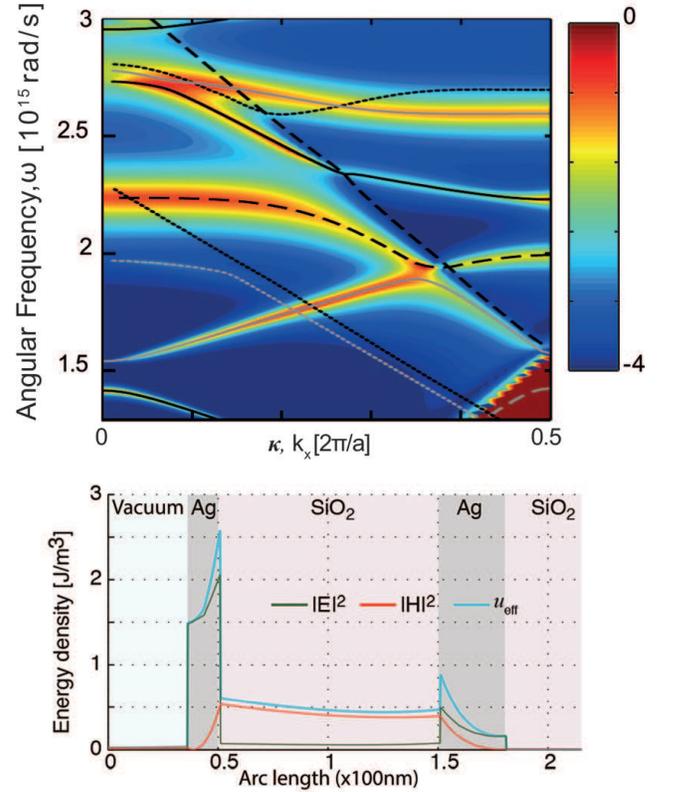}
\caption{Top: Band diagram (lines) overlaid on colored extinction plot. Dominant mode character at $\kappa=0$: ISP (solid), ESP (dashed), LSP (dotted) and parity: black (even) and gray (odd). Bottom: Cross-sectional EM energy density, $u_{\textrm{eff}}$ of ISP, with electric and magnetic contributions.}
\label{fig:absorb}
\end{figure}

The extinction, computed as $\mathcal{A}=\log[1-(\mathcal{T}+\mathcal{R})]$, is shown in Fig.~\ref{fig:absorb} (top panel) along with the calculated dispersion relations. The dominant mode character of each band (as found at $\kappa=0$) is shown with the line style, and the mode parity is indicated by line color. The extinction is plotted for a TM polarized plane-wave incident from the vacuum side of the structure at an angle $\theta$ as in Fig.~\ref{fig:gcart}. The reflectance is obtained from discrete integration of the reflected electromagnetic power normalized to the incident field. Correspondingly, $\mathcal{T}$ is calculated via the normalized power flux flowing through the computational domain's bottom output port. The spectral and angular characteristics of the extinction are seen to correlate strongly with the optical modes. The strong extinction about the ISP modes is mostly due to absorption of their tightly confined EM fields, with higher extinction seen for the antisymmetric ISP branches, as expected.\cite{Dionne:2006p3531} As an example, the bottom panel of Fig.~\ref{fig:absorb} shows the cross-sectional time averaged EM energy density of an ISP mode at $k_x=0.22, \omega = 1.76 \times 10^{15}\unitfrac{rad}{s}$. A significant portion of the energy is concentrated within the metal films, indicating large losses for this mode. The cross-sectional plots shown in figures 3 and 4 are taken along the center of the computational domain, parallel to the y-axis. 

The parity of the ISP modes alternates with increasing frequency. Under normal illumination only even parity modes are excited due to the constant phase of the transverse electric field $E_x$. Thus the loss near $k_x = 0$ mostly occurs at the even mode. As the incidence angle increases, the phase variation along the $x$-direction enables direct excitation of the odd parity ISP mode, accompanied by increased coupling into that branch. 
We also note strong ESP-ISP level repulsion near $\omega = 1.9 \times 10^{15}\unitfrac{rad}{s}$ at $k_x\approx0.35$, indicating hybridization of the two modes.\cite{Rohde:2007p415} As we show below, 
excitation of the ISP modes provides the mechanism for large, ultra-narrow signal transmission, strongly pinned to these resonances.

\begin{figure}
\includegraphics[width=85mm]{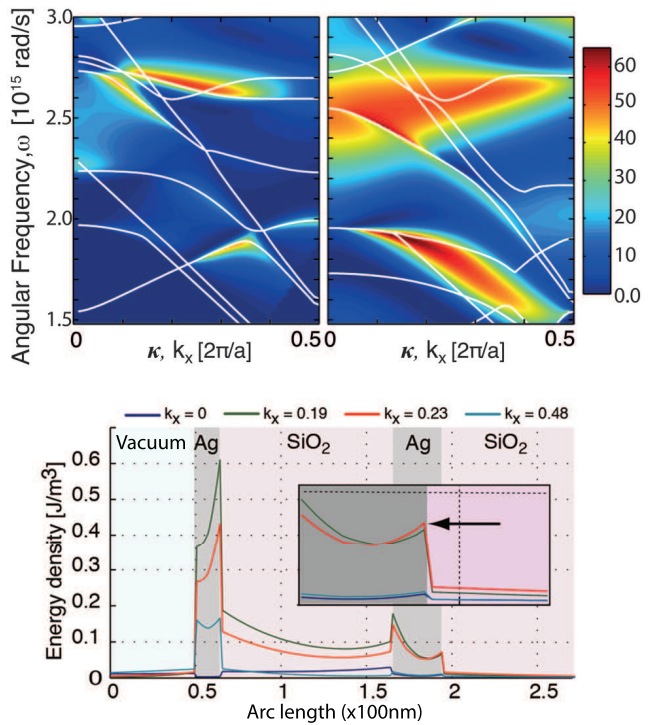}
\caption{Top: Modes overlayed on transmittance ($\%$) contours through: modulated MIM (left) and modulated IM coating obtained as described in text (right). Bottom: EM energy density in MIM, for various incidence angles. Inset: Enlarged view of ISP at innermost silver/silica interface, with arrow indicating maximal outcoupling to SPP for $k_x=0.23$.}
\label{fig:tbd}
\end{figure}

The computed transmittance of the MIM structure is shown in the top left panel of Fig.~\ref{fig:tbd}. A distinguishing characteristic of this system is its narrowband and large transmission, strongly pinned to the ISP bands, where extinction is also maximal. To understand the large transmittance, we plot in the bottom panel of Fig.~\ref{fig:tbd} the EM energy densities near $\omega = 2.67 \times 10^{15}\unitfrac{rad}{s}$, for several values of $k_x$. We find that maximal transmittance occurs when the ISP most efficiently overlaps with the outcoupled SPP excited at the innermost silver/silica interface. This can still occur even in the presence of absorption losses (as is the case for the highly confined ISP mode.) When the ISP does not generate significant coupling between the outermost (vacuum/silver) and innermost interfaces, transmission is strongly suppressed.

For comparison we have also calculated the transmittance of a single metal film modulated by the same cylindrical structure. Patterned metal films are known to yield enhanced transmission, resulting from coupled excitation of SPPs on opposing interfaces.\cite{Gwon:2011p3527,Schroter:1998p3516}. To compare this to MIM, we simulated light transmission through a similar system where the top $15\unit{nm}$ silver film has been omitted (IM film). The top right panel of Fig.~\ref{fig:tbd} shows the transmittance of this structure, overlayed on its computed modes. The only obvious similarity between the two transmittance plots in Fig.~\ref{fig:tbd} is the magnitude of the maximal transmitted intensities, which are near 50$\%$ for the two systems. Otherwise we find that unlike in the MIM, the transmittance of the single modulated metal film exhibits very broad spectral characteristics, and instead of being strongly pinned to a particular mode type is confined to several broad regions between the modes.

We demonstrated efficient coupling of EM plane waves to gap plasmons in a cylindrically-modulated MIM system. Both bandwidth and intensity of the transmitted radiation may be tuned through excitation of ISP modes, which serve to efficiently couple the incident fields to outgoing radiation. The effect is independent of the particular modulation or surface structure, and is reproducible by other periodic MIM geometries.

This work was supported by NSF grants DMR-0804433 and DMR-1105077, with additional support from the University of Auckland's Photonfactory.

%

\end{document}